\documentstyle[12pt,aasms4]{article}

\begin{document}
\newcommand{\be}{\begin{equation}}
\newcommand{\ee}{\end{equation}}

\title{ The Distance to the LMC via the Eclipsing Binary HV 2274}

\author{Cailin~A.~Nelson\altaffilmark{1,2,3}}
\affil{cnelson@igpp.llnl.gov}
\author{Kem~H.~Cook\altaffilmark{1,3}}
\affil{kcook@igpp.llnl.gov}
\author{Piotr~Popowski\altaffilmark{1}}
\affil{popowski@igpp.llnl.gov}
\author{David~R.~Alves\altaffilmark{1,4}}	
\affil{alves@stsci.edu}

{\footnotesize
\altaffiltext{1}{Lawrence Livermore National Laboratory, Livermore, CA 94550}
\altaffiltext{2}{Department of Physics, University of California,
        Berkeley, CA 94720 }
\altaffiltext{3}{Visiting Astronomer, Cerro Tololo
Inter-American Observatory, which is operated by the Association of
Universities for Research in Astronomy, Inc., under cooperative agreement with
the National Science Foundation.}
\altaffiltext{4}{Space Telescope Science Institute, Baltimore, MD 21218}
}

\begin{abstract}

We present $UBV$ photometry of the eclipsing binary Harvard Variable 2274
in the Large Magellanic Cloud (LMC).
The stellar parameters of this binary system were calculated by 
Guinan et al.~(1998a) 
who gave both a reddening towards HV 2274 of $E(B-V)=0.083\pm0.006$
and a distance modulus to the LMC, $\mu_{LMC}=18.42\pm0.007$.  The reddening
of this system was also determined  by Udalski et al.~(1998) who found $E(B-V)=0.149\pm0.015$. With Udalski et al.~(1998) $B$ and $V$ photometry
Guinan et al. (1998b) obtained $E(B-V)=0.12\pm0.009$ and $\mu_{LMC}=18.30\pm 0.07$.
Using our $UBV$ photometry, we derive a reddening of
$E(B-V)=0.088\pm0.025$, consistent with the original value of Guinan et al.
(1998a) and supporting a longer distance modulus to the LMC of 
about $\mu_{LMC}=18.40\pm0.07$. We stress the uncertainties inherent in
ground-based UBV photometry and the concomitant uncertainties in determining
distances based upon such photometry.

Subject Headings: distance scale --- dust, extinction --- Magellanic Clouds 
--- stars: fundamental parameters (colors) --- techniques: photometric
\end{abstract}

\section {Introduction}

The Hubble constant, $H_0$, is one of the most important and heavily
investigated cosmological parameters.  At the moment, the most reliable way to
measure $H_{0}$ is to determine the recession velocity and distance of objects
with motions dominated by the Hubble linear expansion.  Extragalactic
distances are determined by constructing a distance ladder, the first rung of
which is often occupied by the LMC.  Since the uncertainties in the radial
recession velocities are typically very small, the main error in $H_{0}$ comes
from the error in distance.  Therefore, to obtain the true absolute value of
$H_{0}$, one requires the true value of the distance to the LMC.  Different
distance determinations to the LMC result in conflicting values of the
distance modulus spanning the range between $\mu_{LMC}=18.1$ (Stanek, Zaritsky
\& Harris 1998) and $\mu_{LMC}=18.7$ (Feast \& Catchpole 1997), an astounding
spread of 27\% in distance.

One potentially accurate method for determining the distance to the LMC 
is the derivation of stellar parameters from eclipsing binary stars. This
technique dates back to at least 1974 (Dworak 1974).  More recently, Bell
et al.~(1991) and Bell et al.~(1993) determined distances to the Magellanic
Clouds using ground-based spectroscopy and photometry.  Of course, the best
experimental method for determining eclipsing binary stellar parameters would
be acquisition of accurate, space-based, broad spectral range data, e.g.
HST/STIS spectra. The many parameters which must be solved for simultaneously may
give rise to regions of degeneracy in the fit if the spectra do not 
extend over a wide enough range in wavelength.  One way to compensate for 
a narrow spectral range is to include ground-based photometry
in the fit.  Unfortunately, this method also reintroduces all the uncertainties
inherent in ground-based photometry.

The first eclipsing binary used to determine the distance to the LMC 
with space-based spectrophotometry was 
HV 2274, a system whose first CCD light curve was presented
by Watson et al.~(1992), along with estimates of maximum light magnitudes and colors
of $V\approx14.16$ mag and $(B-V)\approx-0.18$ mag.
The stellar parameters of this system were determined by Guinan et al.~(1998a,b). 
By splicing together four HST/FOS 
spectra to form a single spectrum spanning 1150 $\dot{A}$ to 4820 
$\dot{A}$, Guinan et al.~(1998a,b)
used ATLAS 9 model atmospheres (Kurucz 1991) to simultaneously fit for the emergent flux
at the surface of the stars, the reddening, $E(B-V)$,
the normalized interstellar extinction curve, $k(\lambda - V)$, and the ratios
of the stellar radii to the distance of the binary. In their original fit to
the spectra, Guinan et al.~(1998a) included the photometric points of Watson et al.~(1992)
and found a reddening of $E(B-V)=0.083\pm0.006$ mag and a distance 
modulus $\mu_{LMC}=18.42\pm0.07$ mag.  While generally robust, this simultaneous
fitting for both  reddening and extinction admits a possible degeneracy
in the region $E(B-V)\approx0.08-0.12$ mag.  This region arises because similar
relative extinction corrections, $E(B-V)\times k(\lambda-V)$, may be obtained
with varying values of $E(B-V)$ (Guinan et al. 1998b).

The reddening towards HV 2274 was also determined by Udalski et al.~(1998; thereafter U98), who obtained
$UBVI$ photometry of the binary and the surrounding field.  Consistent
with the stellar parameters derived by Guinan et al.~(1998a), U98 found 
that the color of the system remained nearly constant, independent of the 
binary phase.  In agreement with Watson et al.~(1992), U98 found a maximum $V$
magnitude of 14.16 mag.  However, U98 disagreed with the Watson et al~(1992)
$(B-V)$ color, instead finding $(B-V)=-0.129$ mag.  In addition, U98 found  
$(U-B)=-0.905$ mag. 

This multi-color photometric data for HV 2274 and surrounding O/B stars 
allowed for an independent determination of the reddening towards HV 2274. 
By plotting a $(U-B)$ vs $(B-V)$ diagram for early spectral type stars, the
reddening may be determined by  measuring their displacement from a 
sequence of unreddened stars.  Adopting a direction
of reddening  $E(U-B)/E(B-V)=0.76$ (Fitzpatrick 1985), U98 found a reddening
to HV 2274 of $E(B-V)=0.149\pm0.015$ mag. This value was consistent with the 
reddening they found to other $B$ stars in the surrounding field.  Guinan et
al.~(1998b)
then refit their spectra using only the $B$ and $V$ 
photometric points from U98 and found that the multi-parameter
fit swung to the other end of the degeneracy region and gave $E(B-V)=0.12$
mag.
This reduced the distance modulus to the LMC to
$\mu_{LMC}=18.30\pm0.07$ mag, about 2 sigma less than the original.

As we will discuss in detail below, however, the uncertainties in $U$ and $B$
photometry are numerous.  $U$-band photometry is quite sensitive to the specific
filter used for spectral types O and B, exactly those stars used by U98
to determine
the reddening.  On the other hand, $B$-band photometry is sensitive to the
specific filter used for spectral types A and F. The U98
observations were made using a non-standard $U$-band filter with a steeper cutoff at short wavelengths
($\lambda<3500 \dot{A}$) than is typical, and were calibrated using only a few
standard fields containing mostly spectral type A and F standard stars. 
Therefore, we try to verify the reddening results of U98
using a more typical $U$ filter and a larger sample of standard stars.

\section{Observations and Photometric Reductions}

$UBV$ photometry of HV 2274 was carried out on the night of Oct. $18^{th}$, 
1998 at the 0.9-m telescope at Cerro Tololo
Inter-American Observatory (CTIO).  We used the Site2K\_6 CCD at
Cassegrain focus, a $2048\times2048$ CCD with a plate scale of
0.4 arcseconds/pixel and a total field of view of 13.1
arcminutes on a side. The filters used were CTIO VTek2 5438/1026, BTek2
4200/1050 and U\#2 3570/660. The U filter was made of UG1 glass with a 
CuSO4 liquid solution as a red leak blocker. Conditions were photometric.
Table~1 contains the epochs and
durations of 
our observations of HV 2274.  All images were bias subtracted and flat-fielded
in the standard manner.
The $B$-band and $V$-band observations were flat-fielded using dome-flats
corrected by a median of twilight and dawn sky-flats.  The $U$-band observations
were flat-fielded using the median of twilight and dawn sky-flats taken
over four nights.

\begin{figure} 
\centering \leavevmode
\epsfxsize=10cm
\epsffile{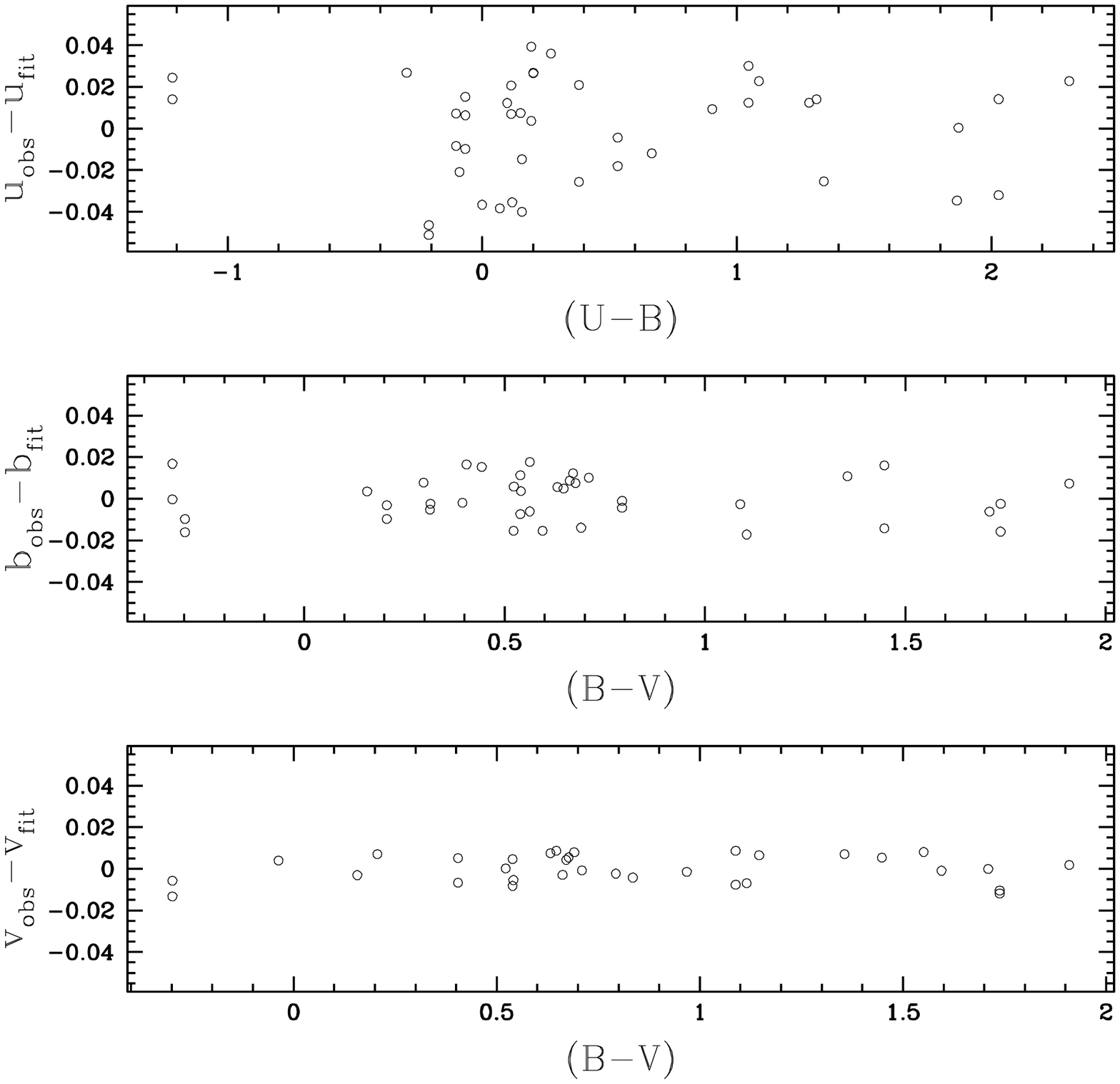}
\caption{The residuals of our transformation versus color, expressed in the 
sense of observed minus fitted magnitude. As is typical for this passband, 
the residuals in U are comparatively large.  However, we see no systematic increase in the size
of residuals in any of our filters as we go to bluer colors.}
\label{residuals}
\end{figure}

\begin{figure} 
\centering \leavevmode
\epsfxsize=10cm
\epsffile{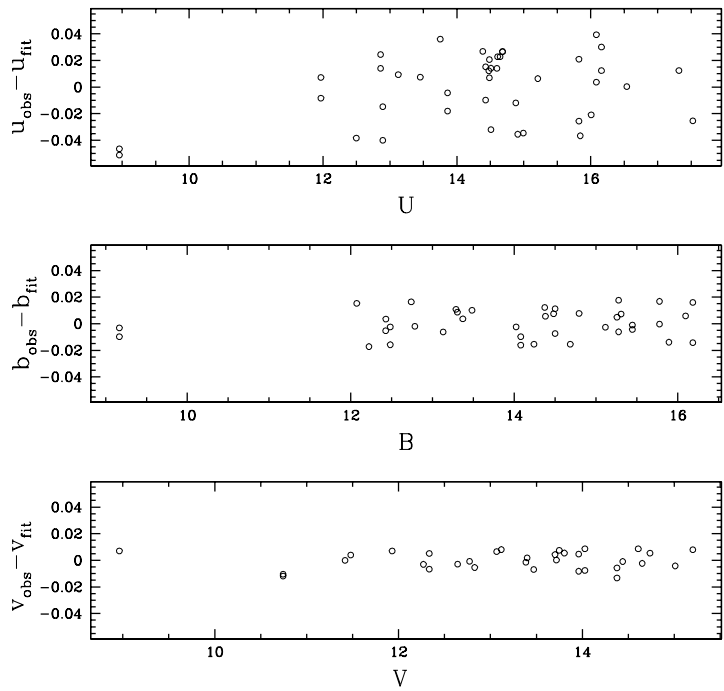}
\caption{The residuals of our transformation plotted versus standard magnitude, expressed in the sense of observed minus fitted magnitude.}
\label{residuals2}
\end{figure}

In order to obtain
transformations from instrumental to standard magnitudes, 
we also observed 51 standard stars from 6 different Landolt
(1992) fields and one E-Region (Graham 1982).  Instrumental
magnitudes were calculated from aperture photometry performed with a radius of 5.8 
arcseconds using DAOPHOT II (Stetson 1987, 1991). 
Using a standard least squares fitting routine, we performed transformations 
of several functional forms, fitting both observed colors and magnitudes as
various functions of standard color, standard magnitude and airmass. 
We found that all fits produced comparable
results that changed the final colors of our O/B stars by at most 0.01
mag.  The adopted magnitude transformations allow for an  
easy comparison on a filter by filter basis with future observations.  In the 
following expressions we represent observed magnitude as the lower case letter,
standard magnitude as the upper case letter and the airmass as $X$.
\be
u = U + 4.734 - 0.090(U-B) + 0.444X
\ee
\be
b = B + 3.125 + 0.107(B-V) + 0.263X
\ee
\be
v = V + 2.974 - 0.021(B-V) + 0.138X
\ee
The range of airmass and color of our
standard star observations was sufficient to include the value of the
atmospheric extinction coefficient as a free parameter in our fit.  The 
atmospheric extinction
coefficients determined for this run are 0.444, 0.263, and 0.138,
for the $U$, $B$ and $V$ bands, respectively.  These
compare well to the values of 0.453, 0.277 and 0.150
calculated by Landolt (1992) as the mean value of 13 years of observing runs
at CTIO.  
In Figure~\ref{residuals}, we show the residuals of our transformations versus standard color
in the sense of observed magnitude minus the fitted magnitude.  We find root
mean square
residuals of 0.024, 0.010 and 0.006 mag for the $U$, $B$ and $V$ passbands respectively.
As is typical for this passband, the residuals in $U$
are comparatively large. However, we see no systematic increase in the size of 
residuals in any of our filters as we go to bluer colors.
In Figure~\ref{residuals2}, we show the residuals of our transformations
versus standard magnitude.  We see no evidence for a dependence of residual
size with magnitude.

Point spread function (PSF) fitting photometry was performed for all the 
stars in
the HV2274 field.  We chose this method instead of aperture photometry since although HV 2274 
is well separated from its neighbors, this was not 
the case for the other O/B stars in the field for which we wish to tabulate
reddenings.  Thus we performed PSF fitting photometry for all the stars
in our program field.   
Both the PSF and the aperture
correction vary across the field of the CTIO 0.9m $2K \times 2K$ CCD.
We used the
photometry package DAOPHOT II which allows for a quadratically varying PSF.
The variation of the aperture correction across the field is only of 
order 0.03 mag. Even small relative shifts, however, can significantly
disturb the morphology of a given set of O/B stars in the color-color plot.
Thus, in order to properly account for
the spatial variation in the aperture corrections, we determined the 
aperture corrections independently for each of the O/B stars.
The aperture correction was determined from a neighbor-subtracted 
image.  For a few of our stars, an imperfect PSF resulted in small regions of 
oversubtraction,
which produced noisy growth curves
and unreliable aperture corrections.  For these stars we adopted the aperture
correction of their nearest bright neighbor.  We note that in the case of HV
2274
aperture photometry and PSF fitting photometry produced identical
results.  

Our results for HV 2274 are,
\be
V=14.203\pm0.006
\ee
\be
(B-V)=-0.172\pm0.013
\label{bmv}
\ee
\be
(U-B)=-0.793\pm0.027
\ee
We include in our error both the formal photon counting errors and an
estimation of the errors in the transformations from instrumental to standard 
magnitude.  We compute the transformation error
as the root mean square of the residuals
shown in Figure~\ref{residuals}.
In Table~2 we compare our results with past efforts.  Since
U98 found the color of the binary to be invariant, we
compare our colors calculated at a specific phase between eclipses
with those of U98
and Watson et al.~(1992) calculated as means across the entire cycle.
Our $V$-band magnitude is close to the maximum light found by U98, 
while our $(B-V)$ color is 
consistent with the original photometry published by Watson (1992).  

\begin{figure}
\centering \leavevmode
\epsfxsize=10cm
\epsffile{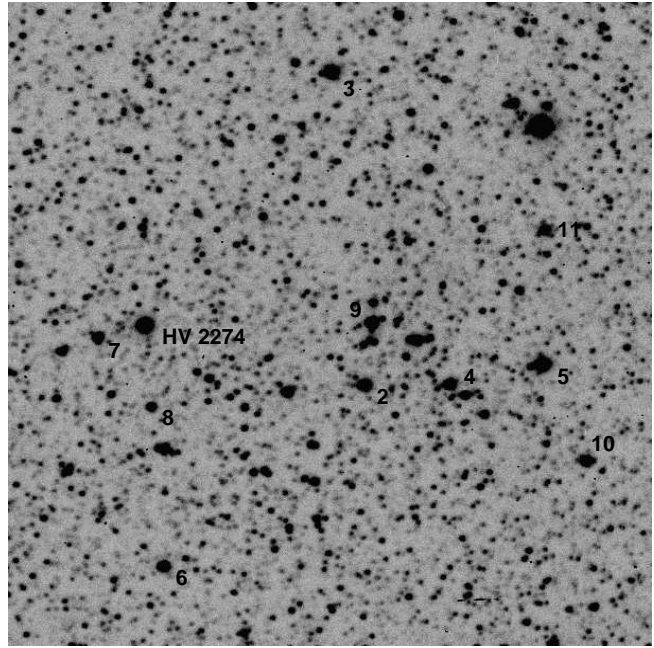}
\caption{A finding chart for HV 2274 and neighboring O/B stars. This image
was taken in the $B$ passband and is oriented such that north is down and 
east is right.}
\label{finding}
\end{figure}

\begin{figure}
\centering \leavevmode
\epsfxsize=10cm
\epsffile{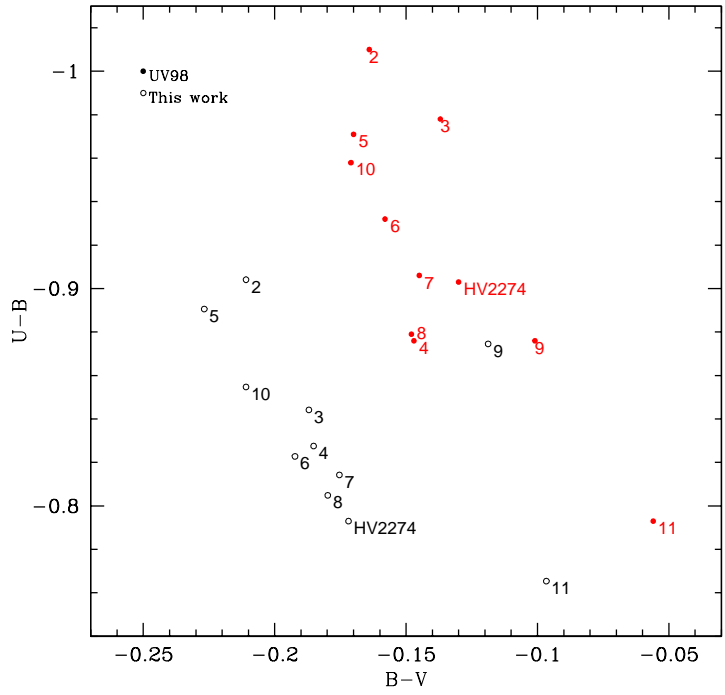}
\caption{A comparison of our photometry with that of Udalski et al.~(1998).
Although the morphology is similar in both sets of photometry, we see a 
significant shift in our photometry towards redder $(U-B)$ and bluer $(B-V)$.}
\label{col_col}
\end{figure}

Table~3 provides a complete listing of our photometry of HV 2274 and 
ten other B stars in the surrounding field, while Figure~\ref{finding} is
a finding chart. These stars are included in the eleven-star
set used by U98 
and for ease of comparison, we adopt the same ID numbers.
We exclude star 1 from our set as it appeared confused 
in our images.  In Figure~\ref{col_col}, we compare our results with that of
U98 in
a $(U-B)$ versus $(B-V)$ plot.  We note that although  
there is a significant shift to redder $(U-B)$ and 
bluer $(B-V)$ in our data, the overall morphology of the set of stars is similar.  

Although we are able to compare our colors with those of U98, 
the observed magnitudes were not published and 
we cannot make comparisons on a filter by filter basis.  This makes
it difficult to determine for certain where the color shifts originate. 
We may, however, speculate on a few possible sources of the problem.

First, we consider the possibility that at least part of the offset is due to the 
steep short-wavelength cutoff of the $U$ filter used by U98.  Differences in short-wavelength
cutoff of the $U$- band can introduce systematic deviations, significant mainly
for B type stars (Bessell 1986). The sign of this effect is
such that a steeper blue cutoff
results in smaller $U$ magnitudes for stars with $(U-B) < 0$ (Bessell 1990).  Assuming 
for a moment agreement in $B$ magnitudes, this implies that spectral type B stars
will appear bluer in $(U-B)$ through a filter with a sharper short-wavelength cutoff. 
We see evidence for this effect in the expected direction as the
$U$ filter used by U98 has a sharper short-wavelength cutoff than does the
CTIO filter, and the U98 results are bluer in $(U-B)$.

However, the magnitude of the difference in the $(U-B)$ photometry
is larger than is expected from solely a $U$
passband mismatch.  Therefore, we also consider the possibility that problems in
the $B$ passband may contribute to the discrepancy in $(U-B)$ as well as the 
discrepancy in $(B-V)$.
Because the $B$ band is situated near the confluence of the Balmer lines, small
shifts in the short wavelength cutoff of the $B$ filter will affect the colors of A and F
stars (Bessell 1990). Many of the standard stars used both by our group and 
U98  
were spectral type A and F stars.  This means that both our
transformations were largely driven by those stars most sensitive to
$B$ passband mismatches.  A small $B$
passband mismatch could therefore introduce significant shifts in the
transformations from observed to standard magnitudes.  However, our complete
spread of standard stars includes stars of spectral types O through M,
including many more red stars than U98.
We deem it unlikely that our transformation could be adversely affected by
difficulties in the A and F stars and yet emerge with a satisfactorily linear 
color term over the entire range of stellar characteristics.

Conversely, $V$-band problems cause significant changes mainly for M stars 
which are much redder than
the majority of standard stars used.  Therefore, the discrepancies in
$(B-V)$ are likely to be due to a $B$-band mismatch, rather then a problem
with the $V$-band.

\begin{figure}
\centering \leavevmode
\epsfxsize=10cm
\epsffile{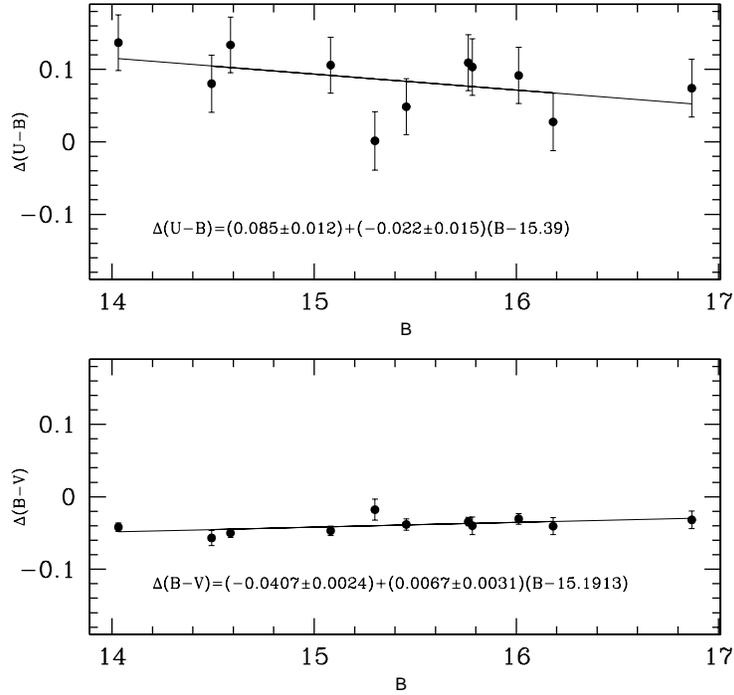}
\caption{A direct comparison of our photometry with that of Udalski 
et al.~(1998).  
The dashed line shows a linear least squares fit to the data with the
corresponding equation recorded below. The errors in $\Delta (B-V)$ include
only our photon counting errors, while we have added an additional error of 
$0.0378$ to $\Delta (U-B)$ in order to produce an acceptable $\chi^{2}$. 
The zero points represent significant offsets between our and U98 photometry
while the slopes are weakly significant and may suggest some non-linearity 
in the transformations from observed to standard magnitudes.}
\label{compare}
\end{figure}

Another possible source of discrepancy is nonlinearities in the
transformations from observed to standard magnitudes.
In Figure~\ref{compare} we again compare our colors to those of U98 
this time by plotting our color residuals versus magnitude, in the sense
$\Delta Color = Color_{\rm This\_work} - Color_{\rm U98}$. 
Most of the errors in $\Delta Color$ are highly correlated between individual 
points and between our and the U98 study. Therefore, as a first approximation,
we assume that the only truly 
uncorrelated errors are the ones associated with the photon counting noise. 
A linear least squares fit to the eleven points using only our estimated
photon noise (U98 had more observations and so their photon noise should be
negligible with respect to ours) yields $\chi^{2}_{\Delta (U-B)} = 151.85$ and 
$\chi^{2}_{\Delta (B-V)} = 7.26$.  
The small $\chi^{2}_{\Delta (B-V)}$ suggests that there is little additional
uncertainty due to, say, different transformations or statistical error.
The linear relation is, therefore, well established and given by 
$\Delta (B-V)= (-0.401 \pm 0.002) + (0.007 \pm 0.003)(B-15.19)$, where the 
errors in the slope and zero point are uncorrelated.
The very high $\chi^{2}_{\Delta (U-B)}$ is a strong evidence that either
$\Delta (U-B)$ versus $B$ relation cannot be represented as a straight line
(indicating severe photometric problems) or that there are additional 
sources of uncorrelated noise, which we did not take into account.
To investigate the second possibility, we renormalize $\chi^2$ to the number
of degrees of freedom, i.e. we fix $\chi^{2}_{\Delta (U-B)}= 9.00$.
This procedure suggests the existence of 0.038 mag of additional error
which should be added in quadrature to each point.
We note that 0.038 is of order of a {\em total} error in $U$-magnitudes 
claimed by both our and U98  groups and can be understood as resulting from
additional statistical uncertainty due to only 4 U98 epochs, systematic errors
from U flat fielding and error in determining the U extinction.
In Figure~5 we show the augmented errors for $\Delta (U-B)$ and our photon 
errors for $\Delta (B-V)$.
The zero points of the linear relations represent an approximate offsets 
between our and the U98 
photometry, but the errors [especially in $\Delta (B-V)$] are smaller than
the actual ones because they were specifically chosen to judge the significance of the 
slopes.  The slope in $\Delta (B-V)$ is significant at the 2.3-$\sigma$ level
and the slope in $\Delta (U-B)$ with renormalized $\chi^2$ is
significant at 1.5-$\sigma$ level. This suggest some
non-linearity in the transformations from observed to standard 
magnitude in either our work or that of U98.

\section{The Reddening to HV 2274}

To determine the reddening to HV 2274 we use the Q parameter method recently
used by Harris, Zaritsky \& Thompson (1997) where Q is defined as

\be
Q = (U-B) -0.76(B-V) -0.05(B-V)^{2}
\ee

This is equivalent to a rotation of a $(U-B)$ versus $(B-V)$ color-color diagram so that
the direction of reddening $E(U-B)/E(B-V)=0.76+0.05(B-V)$ is parallel to the
$(B-V)$ axis.  As detailed in Harris et al.~(1997), the 
coefficient of 0.76 in the direction of reddening was taken
from the ratio of color excesses $E(U-B)/E(B-V)$ evaluated from the average
LMC extinction curve outside 30 Dor (Fitzpatrick 1985), while the coefficient
of 0.05
was derived directly from Galactic studies (Hiltner \& Johnson 1956). 

\begin{figure}
\centering \leavevmode
\epsfxsize=10cm
\epsffile{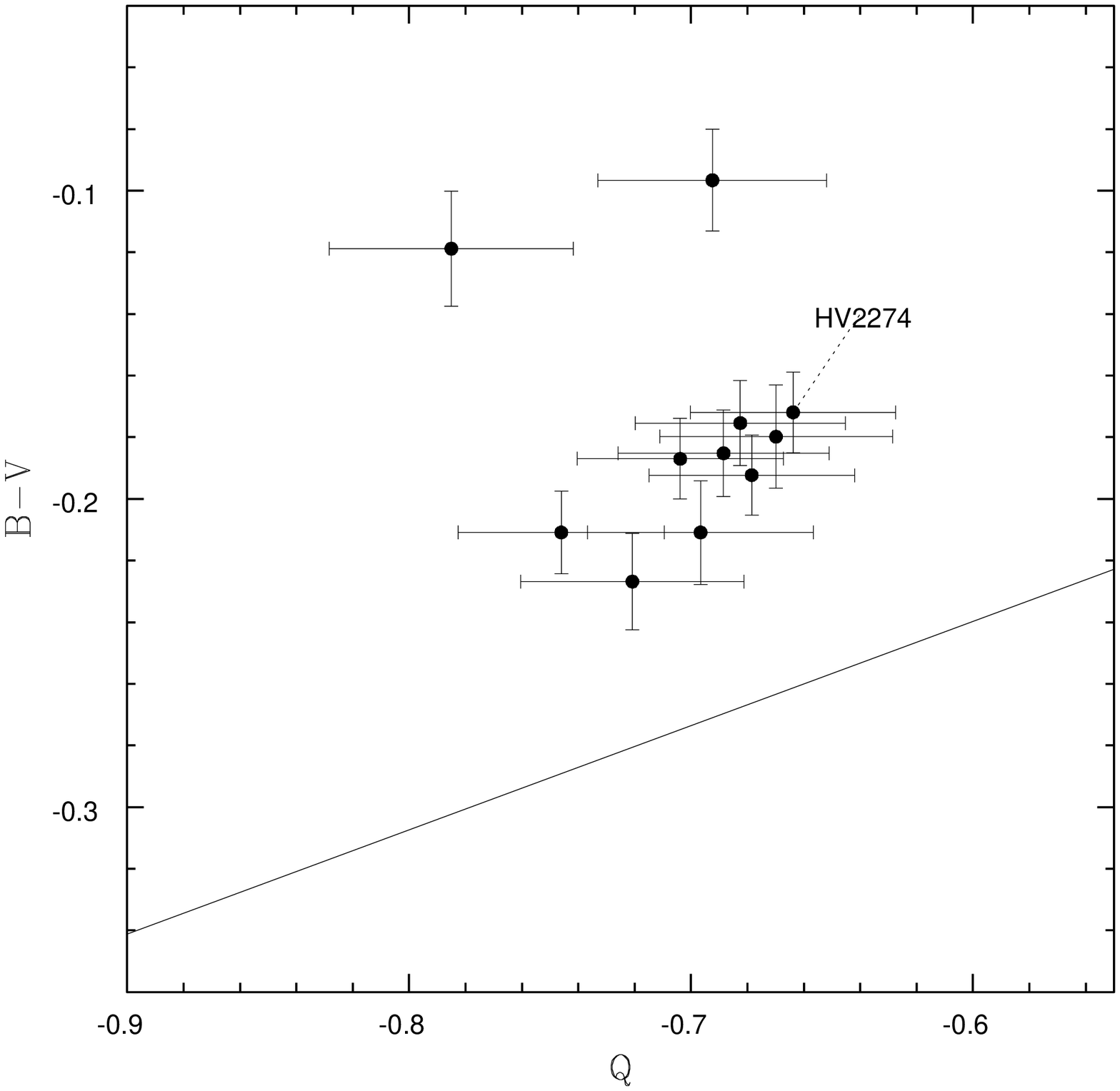}
\caption{$(B-V)$ versus $Q$ plot for HV 2274 and nearby O/B stars. The solid 
line 
is a fiducial of unreddened Galactic O/B stars. The reddening,
$E(B-V)$, of 
a star is measured as its vertical displacement from the unreddened line.}
\label{harris}
\end{figure}

In Figure~\ref{harris} we 
plot a line of unreddened stars and our program stars in a $(B-V)$ versus Q diagram.
Again, our errors include the estimated error in our transformations.
The line of unreddened stars was drawn from Harris et al.~(1997) and was determined by a fitting 
a line in a $(B-V)$ versus Q diagram to a set of observations of unreddened Galactic O/B
stars from Straizys (1992).  The equation of this line is given by
\be
(B-V)=0.338\times Q - 0.036
\ee
The reddening of a program star in such a diagram is then given by the vertical distance
between that star and the unreddened line. We
note that our stars fall in a fairly tight
distribution, parallel to the unreddened line.  We find a median reddening of 
$E(B-V)=0.083$ mag, while for HV 2274 we find 
\be
\label{reddening}
E(B-V)=0.088\pm0.025 
\ee
In Table~4 we
compare this result with previous measurements.
It is noteworthy that this value of $E(B-V)$ derived
from $UBV$ photometry is entirely consistent with the reddening fit by Guinan
et al.~(1998a) using spectra and Watson (1992) $B$ and $V$ photometry.
It is encouraging that two such different methods produced virtually identical
results.

This reddening to HV 2274 is consistent with other literature on this subject.
Unfortunately, HV 2274 lies outside both the LMC reddening maps of Oestreicher
\& Schmidt-Kaler
(1996) and Harris et al.~(1997) and the LMC foreground reddening map of Oestreicher, Gocherman
\& Schmidt-Kaler
(1995).  However, we may still make use of their results to estimate the
likelihood of obtaining a reddening this low.
Oestricher et al.~(1995) finds that the foreground reddening is extremely
patchy and varies from $E(B-V)_{fg}=0.00$ mag to $E(B-V)_{fg}=0.15$ mag, with a mean of
$E(B-V)_{fg}=0.06\pm0.02$ mag.  However, the shape of the frequency 
distribution of foreground reddenings makes quoting a mean value of their distribution
somewhat misleading.  We note that 60\% of their measurements found a 
foreground reddening of less then $0.05$ mag.  Similar conclusions were 
drawn by Bessell (1991), who concurs that the foreground reddening to the LMC
is varied with $E(B-V)_{fg}=0.04$ to $0.09$ mag, and Schlegel, Finkbeiner \& Davis (1998) who
find that the typical foreground reddening measured from dust emission in
surrounding annuli is $E(B-V)_{fg}=0.075$ mag.
Since we are primarily interested in the total reddening to HV 2274,
including reddening within the LMC, we turn again to Harris et al.~(1997) who
present a histogram of total reddening measurements obtained using the Q parameter
method described above. This histogram is drawn from measurements of 2069 O/B
type main sequence stars in a $1.9^{\circ} \times 1.5^{\circ}$ section of the LMC.
Reproducing this histogram and integrating gives us a 15\% probability of obtaining
a total reddening measurement of less then $E(B-V)=0.10$ mag. Therefore, we conclude that  although our derived reddening to HV 2274 is low, it is
not impossibly so. Our reddening is comfortably above the mean foreground reddening and in a
plausible region of total reddening.  

\section{Conclusions}

We obtained $UBV$ photometry of the eclipsing binary HV 2274.
Our principal results are: a reddening, $E(B-V)=0.088\pm0.025$, 
consistent with the original fit of Guinan et al.~(1998a) 
and color, $(B-V)=-0.172\pm0.013$,
consistent with the original photometry of Watson (1992). 
Our results suggest a return to the original estimate of 
$\mu_{LMC} \sim 18.4$ mag.

Even with our generous estimate of error in $E(B-V)$,
our result is still $2 \sigma$ different from that of U98
suggesting a real difference in our photometry.  Consequently, our 
reddening is inconsistent with the later results of Guinan et al.~(1998b), 
emphasizing the sensitivity of the spectrophometric fit to small changes
in the photometry.

Our discussion has emphasized that the systematic uncertainties and variations in 
atmospheric transmission, filters, and calibrations inherent in ground-based data
are likely to limit the accuracy of resulting distances determined from
eclipsing binary systems.  These systematic uncertainties affect distance
moduli at the 5-10\% level and will not be sufficiently reduced with larger 
samples. 
The determination of distances by solving eclipsing binary systems
will become much more reliable with accurate measurements of interstellar
extinction. Ground-based infrared photometry is one way to improve on
the current determinations. The best way, however, would be to extend
space-based spectra over a wide enough range in wavelength to obviate the
need for ground-based spectra altogether.

\acknowledgments

We thank Andrzej Udalski for kindly providing his colors for HV 2274 and
neighbors. We also thank Edward Guinan and Ignasi Ribas for sending us their ephemeris
for HV 2274 in convenient electronic form. Work at LLNL supported by DOE contract W7405-ENG-48.

\clearpage

\clearpage

\begin{deluxetable}{cccc}
\label{table1}
\tablewidth{0pt}
\tablecaption{Summary of Observations}
\tablehead{
\colhead {Filter} & \colhead{Exposure Time (s)} & \colhead {MJD}}
\startdata
U       &       400.0   &       51104.828        \nl
B       &       250.0   &       51104.817        \nl
V       &       200.0   &       51104.820        \nl
\enddata
\end{deluxetable}

\begin{deluxetable}{llll}
\label{table2}
\tablewidth{0pt}
\tablecaption{Photometry of HV 2274}
\tablehead{
\colhead {Author} & \colhead{$V_{max}$} & \colhead {$(B-V)$} & \colhead {$(U-B)$
}}
\startdata
Watson et al.~(1992)  & $14.16\pm0.1$     &    $ -0.18 \pm0.1   $   &           
             \nl
Udalski et al.~(1998) & $14.16\pm0.02$    &    $ -0.129\pm0.015 $   &     $ -0.9
05\pm0.040$    \nl
This work             &                   &    $ -0.172\pm0.013 $   &     $ -0.7
93\pm0.026$   \nl
\enddata
\end{deluxetable}

\begin{deluxetable}{cccc}
\label{table3}
\tablewidth{0pt}
\tablecaption{Photometry of HV 2274 and Surrounding O and B Stars}
\tablehead{
\colhead {Star} & \colhead{V} & \colhead {$(B-V)$} & \colhead {$(U-B)$}}
\startdata
HV2274  &       14.20   &       $-$0.172          &       $-$0.793          \nl
2       &       15.29   &       $-$0.211          &       $-$0.904          \nl
3       &       14.77   &       $-$0.187          &       $-$0.844          \nl
4       &       15.64   &       $-$0.185          &       $-$0.823          \nl
5       &       14.72   &       $-$0.227          &       $-$0.891          \nl
6       &       15.95   &       $-$0.192          &       $-$0.823          \nl
7       &       16.19   &       $-$0.175          &       $-$0.814          \nl
8       &       17.05   &       $-$0.180          &       $-$0.805          \nl
9       &       15.42   &       $-$0.119          &       $-$0.875          \nl
10      &       15.99   &       $-$0.211          &       $-$0.855          \nl
11      &       16.28   &       $-$0.097          &       $-$0.765          \nl
\enddata
\end{deluxetable}

\begin{deluxetable}{lcc}
\label{table4}
\tablewidth{0pt}
\tablecaption{Reddenings to HV 2274}
\tablehead{
\colhead {Author} & \colhead{$E(B-V)$}}
\startdata
Guinan et al.~(1998a)  &   $0.083\pm0.006$  \nl
Udalski et al.~(1998)  &   $0.149\pm0.015$  \nl
Guinan et al.~(1998b)  &   $0.120\pm0.009$  \nl
This work              &   $0.088\pm0.025$  \nl
\enddata
\end{deluxetable}


\clearpage



\begin{references}
\reference{} Bell, S.A., Hilditch, R.W., Reynolds, A.P., Hill, G., \& Clausen, J.V. 1991, \mnras, 250, 119
\reference{} Bell, S.A., Hill, G., Hilditch, R.W., Clausen, J.V., \& Reynolds, A.P. 1993, \mnras, 265, 1047
\reference{} Bessell, M.S. 1986, \pasp, 98, 354
\reference{} Bessell, M.S. 1990, \pasp, 102, 1181
\reference{} Bessell, M.S. 1991, \aap, 242, L17
\reference{} Dworak, T.Z. 1974, Acta Cosmologica, 2, 15
\reference{} Feast, M.W., \& Catchpole, R.M. 1997, \mnras, 286, L1
\reference{} Fitzpatrick, E.L. 1985, \aj, 299, 219
\reference{} Graham, J.A. 1982, \pasp, 94, 244
\reference{} Guinan, E.F., Ribas, I., Fitzpatrick, E.L., \& Pritchard, J.D. 1998a, in Ultraviolet Astrophysics Beyond the IUE Final Archive, eds.
R. Gonzalez-Riestra, R., W. Wamsteker, and R. Harris (ESA SP-413), p. 315
\reference{} Guinan, E.F., Ribas, I., Pritchard, J.D., Bradstreet, D.H., \& Gimenez, A. 1998b, \apj, 509, L21
\reference{} Harris, J., Zaritsky, D. \& Thompson, I. 1997, \aj, 114, 193
\reference{} Hiltner, W.A. \& Johnson, H.L. 1956, \aj, 124, 367
\reference{} Kurucz, R.L. 1991, in Precision Photometry: Astrophysics of the Galaxy, eds. A.G.D. Philip, A.R. Upgren \& K.A. Janes, (Schenectady: L. David Press), p.27
\reference{} Landolt, A. 1992, \aj, 104, 340
\reference{} Oestricher, M.O., Gochermann, J., \& Schmidt-Kaler, T. 1995,
\aaps, 112, 495
\reference{} Oestricher, M.O. \& Schmidt-Kaler,T. 1996, \aaps, 117, 303
\reference{} Schlegel, D.J., Finkbeiner, D.P., \& Davis, M. 1998, \apj, 500, 525
\reference{} Stanek, K.Z., Zaritsky, D., \& Harris, J. 1998, \apj, 500, L141
\reference{} Stetson, P.B. 1987, \pasp, 99, 191
\reference{} Stetson, P.B. 1991, Data Analysis Workshop -3rd ESO/ST-ECF
Garching, 187
\reference{} Straizys, V. 1992, Stellar Photometry, (Tucson: Pachart)
\reference{} Udalski, A., Pietrzy\'{n}ski, G., Wo\'{z}niak, M., Szyma\'{n}ski, M., Kubiak, M. \& \.{Z}ebru\'{n}, K. 1998, \apj, 509, L25 
\reference{} Watson, R.D., West, S.R.D., Tobin, W. \& Gilmore, A.C. 1992, \mnras, 258, 527
\end{references}
\end{document}